\documentclass[12pt]{article}

\usepackage{amssymb, amsmath, latexsym, epsfig, graphicx}

\usepackage[table]{xcolor}
\usepackage{multirow}

\usepackage[round]{natbib} 

\usepackage{bbm, tikz-qtree}

\newtheorem{proposition}{Proposition}[section]

\usepackage{Sweave}
\begin{document}
\Sconcordance{concordance:surv_perm_test_2017_12_11.tex:surv_perm_test_2017_12_11.Rnw:%
1 35 1 1 0 311 1}
\Sconcordance{concordance:surv_perm_test_2017_12_11.tex:./surv_perm_section_sim_2017_12_11.Rnw:ofs 348:%
1 1 1 1 30 1 1 1 4 2 1 1 7 22 1 1 20 17 1 1 4 1 2 12 1 1 21 12 1 4 2 21 %
1 2 2 1 1 2 2 1 1 2 2 1 1 1 2 1 3 23 1 2 2 1 1 2 2 1 1 2 2 18 1 2 2 1 1 %
2 2 1 1 2 2 7 1 1 41 28 1}
\Sconcordance{concordance:surv_perm_test_2017_12_11.tex:surv_perm_test_2017_12_11.Rnw:ofs 552:%
349 6 1}
\Sconcordance{concordance:surv_perm_test_2017_12_11.tex:./ce_example_2017_12_11.Rnw:ofs 559:%
1 1 1 1 10 3 1 1 16 3 1 1 16 1 1 1 19 2 1 2 13 1 1 1 24 1 13 3 1 1 15 %
88 1}
\Sconcordance{concordance:surv_perm_test_2017_12_11.tex:surv_perm_test_2017_12_11.Rnw:ofs 671:%
357 30 1}

\thispagestyle{empty} \baselineskip=28pt

\begin{center}
{\LARGE{\bf  A Permutation Test on Complex Sample Data }}
\end{center}

\baselineskip=12pt


\vskip 2mm
\begin{center}
Daniell Toth 
\\
\vskip 2mm
Bureau of Labor Statistics\footnote{Daniell Toth is Senior Mathematical Statistician, Office of Survey Methods Research, 
Bureau of Labor Statistics, Suite 3950, Washington, DC 20212 (email:toth.daniell@bls.gov)}
\end{center}


\begin{center}
{\Large{\bf Abstract}}
\end{center}
Permutation tests are a distribution free way of performing hypothesis tests. 
These tests rely on the condition that the observed data are exchangeable among the groups being tested under the null hypothesis.
This assumption is easily satisfied for data obtained from a simple random sample or a controlled study after simple adjustments to the data, but there is no general method for adjusting survey data collected using a complex sample design to allow for permutation tests.
In this article, we propose a general method for performing a pseudo-permutation test that accounts for the complex sample design.
The proposed method is not a true permutation test in that the new values do not come from the set of observed values in general, but of an expanded set of values satisfying a random-effects model on the clustered residuals.  
Tests using a simulated population comparing the performance of the proposed method to permutation tests that ignore the sample design demonstrate that it is necessary to account for certain design features in order to obtain reasonable $p$-value estimates.

\baselineskip=12pt 


\baselineskip=12pt
\par\vfill\noindent
\par\medskip\noindent
\clearpage\pagebreak\newpage \pagenumbering{arabic}
\baselineskip=24pt

\section{Introduction}

The permutation test is a simple test to assess the significance of association between a random variable and group membership, proposed originally for data from designed experiments \citep{fisher35} and then more generally for observed data \citep{pitman38}.
Given a dataset containing observations of a variable $\mathbf{y} = (y_1, \ldots, y_n)$ and corresponding group (or treatment) labels  $\mathbf{g} = (g_1, \ldots, g_n)$, the permutation test provides an estimate of the distribution of a test statistic, conditioned on the observed data, under the null hypothesis that the group labels are independent of the $y$ values. 
This estimated conditional distribution is constructed by calculating the test statistic for all possible permutations of the observed values under the null hypothesis.  Then a $p$-value is obtained by comparing it to the original value.

Despite being first purposed for designed experiments, which are strongly related to survey sample designs \citep{fienberg96}, 
permutation tests have not been generally applied to survey data as they have for experimental design data \citep{good05}. 
Indeed, the key assumption of exchangeability \citep{kingman78} is often violated for survey data. 
Unlike experimental designs where simple adaptions to the test statistic have allowed for these tests to be applied to the data, there have been no adaptions purposed that allow for permutation tests to be applied to data collected using a general complex sample design.
 
The purpose of this article is to propose a method, following the procedure of \cite{welch90}, for a randomization test on group effects using data obtained from a complex sample.
In order to permute values within and across clusters, we adopt a model based method like that of \cite{scott82} for estimating cluster effects, leading to what we call a pseudo-permutation test.  
We show that estimating this model does not prevent the method from leading to a consistent permutation test under certain conditions. 
In Section \ref{sec:ptest} we describe a general permutation test on independent, identically distributed (iid) data and then provide a method for conducting the test on complex sample data.
We demonstrate the method through simulations in Section \ref{sec:sims}.
In Section \ref{sec:app} we apply this method to an analysis of consumer expenditure data.
A discussion of the results is provided in Section \ref{sec:disc}.

\section{Permutation Tests}\label{sec:ptest}

Consider a data set consisting of $n$ observations $\mathbf{y} = (y_1, \ldots, y_n)$ of a continuous random variable $Y,$ along with corresponding group labels $\mathbf{g} = (g_1, \ldots, g_n)$ from the random variable $G$. 
Permutation tests are based on the idea that if $Y$ is independent of the group labels $G$, then we are as equally as likely to have observed a dataset with the same observed values $\mathbf{y}$ and $\mathbf{g},$ but with the assignment between the values and the group labels permuted.
A test statistic is computed on several permuted datasets and compared to the value of the test statistic under the observed order.
If the value of the observed test statistic is considered too extreme based on the values over several permutations, the null hypothesis of independence is rejected.

\subsection{A Test on iid Data}\label{sec:iid}

Suppose $Y$ is a continuous random variable satisfying
\begin{equation}\label{iid_mod}
Y_i = \mu_{g_i} +\epsilon_i,
\end{equation}
for $i = 1 \ldots n,$ where  $\mu_{g_i}$ are unknown constants and each $\epsilon_i$ is an independent and identically distributed (iid) random variable with mean 0 and finite variance from an unknown density $f_{\epsilon}.$
Without loss of generality, we will only consider the case when there are two groups $g_i \in \{0, \, 1\}$ and we are testing the hypothesis
\begin{equation}\label{mean}
E[Y \; | \; G=0] = E[Y \; | \; G=1],
\end{equation}
or equivalently $\mu_0 = \mu_1.$
The conditional probability of observing $\mathbf{y}$ given $\mathbf{g}$ is
\begin{equation*}\label{like}
P\big(\mathbf{Y} = \mathbf{y} \; | \; \mathbf{G} = \mathbf{g}\big) = \prod_{i=1}^n f_{\epsilon}(y_i -\mu_{g_i} \; | \; g_i).
\end{equation*}
If $\mu_0 = \mu_1 = \mu,$ then $\mathbf{y}$ is independent of $\mathbf{g},$ so the probability of $\mathbf{y}$ given $\mathbf{g}$ becomes
\begin{equation*}\label{iid}
P(\mathbf{y} \; | \; \mathbf{g}) = \prod_{i=1}^n f_{\epsilon}(y_i -\mu_{g_i}).
\end{equation*}

Given a vector $\mathbf{v},$ let $\sigma(\mathbf{v})$ represent a random permutation of $\mathbf{v}$ and if $v_i$ is the $i$th value in $\mathbf{v}$, let $v_{\sigma(i)}$ denote the $i$th value in the permuted vector $\sigma(\mathbf{v})$.
Then, since $\epsilon_1, \ldots, \epsilon_n$ are iid, the distribution of $\mathbf{y}$ given $\mathbf{g}$ is same as the distribution of the permuted values of $\mathbf{y},$  $\sigma(\mathbf{y})$ under the null hypothesis \citep[Chapter~6.2]{cox79}. 
That is
\begin{equation}\label{iperm}
P(\mathbf{y} \; | \; \mathbf{g}) = \prod_{i=1}^n f_{\epsilon}(y_i -\mu_{g_i}) = \prod_{i=1}^n f_{\epsilon}(\epsilon_i) = \prod_{i=1}^n f_{\epsilon}(\epsilon_{\sigma(i)}).
\end{equation}
Therefore, we can estimate the conditional distribution of any finite, deterministic function of $(\mathbf{y}, \mathbf{g}),$ $\psi = \psi(\mathbf{y}, \mathbf{g}),$ by computing the values of $\psi$ using permutations of the values of $\mathbf{y}.$  
These values provide an empirical distribution that is conditional on the observed values of $\mathbf{g}$ under the assumption that $\mathbf{y}$ and $\mathbf{g}$ are independent.

For example, in order to test the hypothesis given in (\ref{mean}) we consider the test statistic 
\begin{equation*}\label{meandiff}
\psi = n_{1}^{-1} \sum_{i=1}^{n} y_i \mathbbm{1}_{\{g_i=1 \}} - n_{0}^{-1} \sum_{i=1}^{n} y_i \mathbbm{1}_{\{g_i=0 \}} = \hat{\mu}_1 - \hat{\mu}_0,
\end{equation*}
where $n_{g} = \sum_{i=1}^{n} \mathbbm{1}_{\{g_i=g \}}$ and $\mathbbm{1}_{\{ \}}$ is an indicator function.
Compute, $\psi_{\mathbf{y}},$ the statistic under the observed order of $\mathbf{y}$ and compare this value to the empirical distribution of
$\psi_{\sigma(\mathbf{y})},$ obtained by computing the test statistic for many random permutations of the values.
If all possible permutations are used to compute the distribution of the statistic, this is called an exact test, whereas if a large number of randomly generated permutations are used to approximate this distribution then it is called a randomization test \citep{good05}.


\subsection{Data from a Complex Sample Design}

Survey data are often collected under a sample design that invalidates the iid assumption for observed units.
A design which causes the distribution of observed values of the variable of interest to be different than the distribution of the variable of interest in the population is called an informative design.
Analysis ignoring the sample design can lead to invalid inference \citep{holt80, pfeffermann93}.

For informative sample designs, auxiliary data must be available for each observation before the sample is drawn for use in the sample design.
These auxiliary data can be used to stratify the population and select observations from each stratum separately, identify clusters to select instead of individual observations, select certain observations with higher probability than others based on the values of an auxiliary variable, or a combination of these.
Some of these design features are likely to provide observed data that violate the assumptions under model (\ref{iid_mod}).
For instance data that are collected from a cluster sample are likely to have observations with values that are more homogeneous within clusters than over the whole population and values of the variable of interest are usually related to the variables used to stratify the population as well as the sample probabilities.

As in \cite{scott82}, we consider a sample of $n$ total observations drawn from $C \leq n$ clusters. 
The observations in the sample are indexed by $\big\{(Y_{i j}, \mathbf{X}_{ij}, G_{i j}) \; | \; j= 1 \ldots n_i \big\}_{i=1}^C$ where $n_i$ is the number of units in cluster $i$ and $n = \sum_{i=1}^{C}n_i.$ 
Here $Y_{i j}$ represents the variable of interest, $\mathbf{X}_{i j}$ is the vector of $p$ auxiliary random variables associated with each unit, and $G_{ij}$ the corresponding group label.
We assume the first variable in the auxiliary data is $X_{ij1}=1$ for each $i$ and $j,$ and that the values of the auxiliary data $\mathbf{X}$ are known for all units in the population.
The values of $\mathbf{Y}$ and $\mathbf{G}$ are collected from the sample units and so only values from the sampled units are observed.
We are interested in testing the null hypothesis
\begin{equation}\label{condmean}
E[Y \; | \; \mathbf{X} = \mathbf{x}, G=0] = E[Y \; | \; \mathbf{X} = \mathbf{x}, G=1],
\end{equation}
using the observed data.

Suppose we model the conditional expectation of $Y$ by the linear equation 
\begin{equation}\label{lin_mod}
E[Y \; | \; \mathbf{X} = \mathbf{x}] = \mathbf{x}\mathbf{\beta}, 
\end{equation}
for some unknown vector of coefficients $\mathbf{\beta}=(\beta_0, \beta_1, \ldots, \beta_p)^T.$
Then the estimated vector of coefficients $\hat{\mathbf{\beta}},$ obtained using the design-consistent estimating equation \citep{binder83}
\begin{equation*}
 \hat{\mathbf{\beta}}= \min_{\mathbf{\beta}} \sum_{i=1}^{C}\sum_{j=1}^{n_i} w_{ij} \big( y_{ij} - \mathbf{x}_{ij}\mathbf{\beta}\big)^2, 
\end{equation*}
 is the solution to the equation
\begin{equation}\label{estequ}
\sum_{i=1}^{C}\sum_{j=1}^{n_i} w_{ij} \big( y_{ij} - \mathbf{x}_{ij}\mathbf{\beta}\big)\mathbf{x}_{ijk}=0,
\end{equation}
for all $k=1 \ldots p,$ where $w_{ij}$ is the sample weight for the observation $j$ in cluster $i.$ 

Define $r_{ij} =  y_{ij} - \mathbf{x}_{ij}\hat{\mathbf{\beta}},$ the residual of the estimated conditional expectation for observation $ij.$ 
Then
\begin{equation*}
 \sum_{i=1}^{C}\sum_{j=1}^{n_i} w_{ij} r_{ij} x_{ijk} = \sum_{i=1}^{C}\sum_{j=1}^{n_i} w_{ij} r_{ij} x_{ijk} \mathbbm{1}_{\{g_{ij}=0 \}} + \sum_{i=1}^{C}\sum_{j=1}^{n_i} w_{ij} r_{ij} x_{ijk}  \mathbbm{1}_{\{g_{ij}=1 \}} = 0,
\end{equation*}
for each $k =  1 \ldots p,$ under the null hypothesis (\ref{condmean}), including $k=1,$ where $x_{ij1}=1.$
We will now derive a permutation test on these residuals as proposed by \cite{gail88}.

Consider the sum of weighted residuals for only units with a particular group label, such as
\begin{equation}\label{teststat}
\psi=\psi(\mathbf{wr}, \mathbf{g}) =   \sum_{i=1}^{C}\sum_{j=1}^{n_i} w_{ij} r_{ij}  \mathbbm{1}_{\{g_{ij}=1 \}}.
\end{equation}
Under the null hypothesis given by (\ref{condmean}), the test statistic defined by equation (\ref{teststat}) has expected value $E[\psi] = 0.$

In order to test the null hypothesis, we need to compute $\psi$ over all permutations of observed values $\big\{w_{ij} r_{ij} \; | \; j= 1 \ldots n_i \big\}_{i=1}^C,$ but unlike the model defined by equation (\ref{iid_mod}) the values of $w_{ij} r_{ij}$ are not necessarily exchangeable.
Though we are accounting for much of the design through the model (\ref{lin_mod}) and the sample weights, values of $w_i r_i$ from different clusters are not necessarily exchangeable.
Therefore, we assume a model for $w_{i j} r_{i j}$ like the one used by \cite{scott69} for multistage surveys,
\begin{equation}\label{clus_mod}
w_{i j} r_{i j} = \eta_{i j} = \nu_i + \epsilon_{ij},
\end{equation}
 where $\{\nu_i \; | \; i= 1 \ldots C\}$ and $\{ \epsilon_{i j} \; | \; i= 1 \ldots C; j = 1 \ldots n_i\}$ are independent, continuous random variables with mean 0 with distribution functions $F_{\nu}$ and $F_{\epsilon_i},$ respectively.
Under these assumptions and the null hypothesis
\begin{align}
P\big(\mathbf{\eta} \; | \; \mathbf{g}\big)
=& \prod_{i}\prod_{j}P(\nu_i + \epsilon_{ij}) P(g_{ij}) \\
=& \prod_{i}f_{\nu}(\nu_i)\prod_{j}f_{\epsilon_i}(\epsilon_{ij}) P(g_{ij})\\
=& \prod_{i}f_{\nu}(\nu_{\sigma(i)})\prod_{j}f_{\epsilon_i}(\epsilon_{i\sigma(j)})P(g_{ij}).\label{nuperm}
\end{align}
This leads to a method for conducting permutation tests using data from a complex sample design by permuting the estimated values of the cluster effects $\mathbf{\nu}$ and error terms $\mathbf{\epsilon}$ according to equation (\ref{nuperm}).
Next we describe a multi-step procedure for obtaining a set of permuted "pseudo"-values of the set $\mathbf{\eta} = \{\eta_{i j} \; | \; i= 1 \ldots C; \; j = 1 \ldots n_i\}.$ 

\subsection{Method and Conditions}

By a random permutation of $\mathbf{\eta},$ we really mean a vector of $C+1$ random permutations $(\sigma_0, \sigma_1, \ldots \sigma_C),$ where $\sigma_0$ is a random permutation of the set of indices $\{1, \ldots, C\},$ and $\sigma_i$ is a random permutation of $\{1, \ldots, n_i\},$ for each $i=1, \ldots, C.$ 
If we denote a set of permuted values of $\mathbf{\eta}$ by $\sigma(\mathbf{\eta}),$ then $\sigma(\mathbf{\eta})=\{\nu_{\sigma_0(i)} +\epsilon_{i \sigma_i(j)} \; | \; i= 1 \ldots C, j = 1 \ldots n_i\}.$ 
The cluster effects $\{\nu_i \; | \; i = 1 \ldots C \}$ are permuted and then the $\{\epsilon_{i j} \; | \; j = 1 \ldots n_i\}$ are permuted within each cluster.
For a given set of observed $\mathbf{\eta},$ let $\mathcal{S}$ denote the set of all such permutations. 

If we randomly select $m>>0$ random permutations from $\mathcal{S}$, 
then for a constant $\alpha >0$ the probability $P\big(|\psi(\mathbf{\eta}, \mathbf{g})| \geq \alpha \big)$ under the null hypothesis 
can be estimated by
\begin{equation}\label{randomtest}
\hat{P}\big(|\psi(\mathbf{\eta}, \mathbf{g})| \geq \alpha \big) = m^{-1}\sum_{\sigma \in \mathcal{S}} \mathbbm{1}_{\{|\psi(\sigma(\mathbf{\eta}), \mathbf{g})| \geq \alpha \}} 
\end{equation}
 \citep[Chapter~6.7]{flury97}.
Note that even though $F_{\nu}$ and $F_{\epsilon_i}$ for $i = 1 \ldots C$ are distribution functions of continuous random variables, the value of the test statistic could be equal, $\psi(\sigma(\mathbf{\eta}), \mathbf{g}) = \psi(\sigma'(\mathbf{\eta}), \mathbf{g}),$ for two different permutations $\sigma$ and $\sigma'$ in $\mathcal{S}.$ 
This occurs if the permuted values of $\{\epsilon_{i j} \; | \; j = 1 \ldots n_i\}$ that have group label $g=1$ are the same under both permutations and $n_{\sigma(i)}(g) = n_{\sigma'(i)}(g)$  for all $i=1, \ldots, C,$ where $n_i(g)= \sum_{j=1}^{n_i} \mathbbm{1}_{\{g_{ij}=1 \}}$ is the number of observations from cluster $i$ that have group label 1.
The unique values of the test statistic applied to permuted $\mathbf{\eta}$ form an equivalence class of permutations in $\mathcal{S};$
let $\mathcal{V}$ be the set of unique values.
Therefore, equation (\ref{randomtest}) is estimating the proportion of permutations that are in equivalence classes with values of the test statistic that are greater or equal to the value of the test statistic on the observed data.

Since the values of $\nu_i$ and $\epsilon_{ij}$ are unknown for each $\eta_{ij} = \nu_i + \epsilon_{ij},$ the next step is to estimate these values in order to perform the permutations. 
The cluster mean, $\nu_i,$ for each cluster $i$ is estimated by 
\begin{equation}\label{cluseff}
\hat{\nu}_i = n^{-1}_i \sum_{j=1}^{n_i} \eta_{i j}= \nu_i + n^{-1}_i \sum_{j=1}^{n_i} \epsilon_{ij},
\end{equation}
and $\hat{\epsilon}_{ij} = \eta_{ij} - \hat{\nu}_i.$
The permuted pseudo-values are obtained by adding the $i$ estimated value of the permuted cluster effects $\hat{\nu}_{\sigma_0(i)}$ to the permuted values of $\hat{\epsilon}_{i \sigma_i(j)}$ in cluster $i.$ 
Since these new values of $\sigma(\eta)_{ij} = \hat{\nu}_{\sigma_0(i)} + \hat{\epsilon}_{i \sigma_i(j)},$ lead to values that are not in the original vector of values $\mathbf{\eta},$ this is not a true permutation of $\mathbf{\eta},$ but rather to a set of pseudo-values.
This set of pseudo-values are the permuted values of $\mathbf{\eta}$ under the assumed model (\ref{clus_mod}) for the true values of $\mathbf{\nu}$ and $\mathbf{\epsilon}.$

The following result states that the effect of replacing the true $\mathbf{\eta}$ with these estimated values $\hat{\mathbf{\eta}}$ in equation (\ref{randomtest}) is small and vanishes asymptotically under certain conditions.
In order to obtain asymptotic results, we consider samples of increasing size, $n,$ from a clustered super-population model satisfying equations (\ref{lin_mod}) and (\ref{clus_mod}).
We use the notation $C_n,$ $\mathcal{V}_n,$ and $\mathcal{S}_n$ to remind us that the number of clusters, unique values of the test statistic, and the set of all possible permutations on the data under the proposed method depends on the sample.
Obviously the data $(\sigma(\hat{\mathbf{\eta}}), \mathbf{g})$ depends  on the sample and sample size but we suppress the subscript $n$ to reduce the complexity of the notation.
The conditions stated for the next result are assumed to occur with probability~1 with respect to this super-population model. 

\begin{proposition}
Suppose a sample of $n$ observations from $C_n$ clusters, $n=\sum_{i=1}^{C_n} n_i,$ is drawn from the super-population model.
If the following conditions are satisfied:
\begin{enumerate}
  \item \label{cond:fvar} $\int^{\infty}_{-\infty} u^2 dF_{\epsilon_i}(u) < M < \infty$ for some $0 \leq M < \infty$ and all $i = 1 \ldots C;$
  \item \label{cond:ediff}  $\exists \delta >0$ such that $\min_{\{x \neq y \; | \; x, y \in \mathcal{V}_n\}} \big|x - y| > \delta$ $\forall n;$
  \item \label{cond:ni}  $\limsup_{n \rightarrow \infty} \; \max_{\{i = 1 \ldots C_n\}} n_i^{-1/2} = o(C^{-1});$
  \item \label{cond:ndiff} $\limsup_{n \rightarrow \infty} \; \max_{\{k \neq l \; | \; k, l = 1, \ldots, C_n\}} |n_k(g) - n_l(g)| = O(1)$
\end{enumerate}
then for $\alpha=\big|\psi(\hat{\mathbf{\eta}}, \mathbf{g})\big|,$
$$\lim_{n \rightarrow \infty} \hat{P}\big(\big|\psi(\sigma(\hat{\mathbf{\eta}}), \mathbf{g})\big| > \alpha \big) = \hat{P}\big(\big|\psi(\sigma(\mathbf{\eta}), \mathbf{g})\big| > \alpha \big).$$
\end{proposition}

\vspace{3pt}

The first condition, Condition \ref{cond:fvar}, assumes the residuals from the model, equation (\ref{lin_mod}), have a finite variance, therefore the Central Limit Theorem applies to the error term obtained from estimating the cluster effect values $\mathbf{\nu}.$ 
Condition \ref{cond:ediff} requires the difference of the absolute values of the test statistic between equivalence classes to be uniformly bounded above 0.
The next two conditions pertain to the sample design.
Condition \ref{cond:ni} requires that the number of observations within each cluster increases as $n$ increases, but allows for the number of clusters $C_n$ sampled to increase as the sample size $n$ increases. 
Condition \ref{cond:ndiff} requires that the difference in the number of observations from a cluster that have group label~1 is bounded for all clusters.

\vspace{3pt}

\emph{{\large proof:}}

Let $\xi_i = n_{i}^{-1} \sum_{j=1}^{n_i} \epsilon_{ij}$ be the error in estimation of $\hat{\nu}_i$ from equation (\ref{cluseff}), then the test statistic defined in (\ref{teststat}) using these permuted pseudo values,
\begin{eqnarray*}
\psi(\sigma(\hat{\mathbf{\eta}}), \mathbf{g}) &=&\sum_{i=1}^{C_n}\sum_{j=1}^{n_i} \big(\hat{\nu}_{\sigma(i)} + \hat{\epsilon}_{i\sigma(j)} \big)  \mathbbm{1}_{\{g_{ij}=1 \}}\\
&=& \sum_{i=1}^{C_n}\sum_{j=1}^{n_i} \big(\nu_{\sigma(i)} + \xi_{\sigma(i)} + \epsilon_{i\sigma(j)} - \xi_i \big)  \mathbbm{1}_{\{g_{ij}=1 \}}\\
&=& \psi(\sigma(\mathbf{\eta}), \mathbf{g}) + \sum_{i=1}^{C_n}\sum_{j=1}^{n_i} \big(\xi_{\sigma(i)}- \xi_i \big)\mathbbm{1}_{\{g_{ij}=1 \}}\\
&=& \psi(\sigma(\mathbf{\eta}), \mathbf{g}) + R,
\end{eqnarray*}
where $R= \sum_{i=1}^{C_n} n_{i}(g)(\xi_{\sigma(i)}-\xi_i).$

Define the inverse function of a permutation $\sigma$ as the integer valued function $\sigma^{-1},$ such that $\sigma^{-1}(i)=k$ implies $\sigma(k)=i.$
Then the difference between the value of the test statistic for the permuted pseudo-values and the true permuted values $R$ can be written 
$R= \sum_{i=1}^{C_n} (n_{i}(g) - n_{\sigma^{-1}(i)}(g))\xi_i.$

By Condition \ref{cond:ndiff} there exists a $0<K<\infty$ such that $R \leq K \sum_{i=1}^{C_n} \xi_i$ for all $n.$
Since the random variable $\xi_i$ is the sum of $n_i$ iid random variables with zero mean and finite variance for each $i = 1, \ldots, C_n,$ by Condition \ref{cond:fvar}, $R$ is a mean-zero random variable with variance $\leq n_{*}^{-1/2}C_nK,$ where $n_{*} = \min_{\{i = 1 \ldots C_n\}} n_i.$
Therefore, by Condition \ref{cond:ni}, $R \rightarrow 0$ as $n \rightarrow \infty$ with probability~1 with respect to the super population model.

Now, let $\alpha$ be the absolute value of the test statistic on the original order of the data, $\alpha=\big|\psi(\hat{\mathbf{\eta}}, \mathbf{g})\big|$ and $\sigma \in  \mathcal{S}_n$ be a fixed permutation of the data $\eta$ using the above procedure.
We now consider the value of $\mathbbm{1}_{\{|\psi(\sigma(\mathbf{\eta}), \mathbf{g})| \geq \alpha \}}$.
If $\sigma$ is in the same equivalence class as the null-permutation, then 
\begin{equation}\label{I}
\mathbbm{1}_{\{|\psi(\hat{\mathbf{\eta}}), \mathbf{g})| \geq \alpha \}} = \mathbbm{1}_{\{|\psi(\sigma(\hat{\mathbf{\eta}}), \mathbf{g})| \geq \alpha \}} = \mathbbm{1}_{\{|\psi(\mathbf{\eta}, \mathbf{g})| \geq \alpha \}} = 1,
\end{equation}
otherwise, 
$\big| \psi(\sigma(\hat{\mathbf{\eta}}), \mathbf{g}) \big| = \big| \psi(\sigma(\mathbf{\eta}), \mathbf{g}) + R \big|.$

Since $R \rightarrow 0$ as $n \rightarrow \infty,$ for large enough $n,$ $R < \delta < \big| \psi(\sigma(\mathbf{\eta}), \mathbf{g}) - \alpha \big|,$ by Condition \ref{cond:ediff}. 
Therefore, 
\begin{equation}\label{II}
\mathbbm{1}_{\{|\psi(\sigma(\hat{\mathbf{\eta}}), \mathbf{g})| \geq \alpha \}} = \mathbbm{1}_{\{|\psi(\sigma(\mathbf{\eta}), \mathbf{g}) + R| \geq \alpha \}} = \mathbbm{1}_{\{|\psi(\mathbf{\eta}, \mathbf{g})| \geq \alpha \}}.
\end{equation}

From equations (\ref{I}) and (\ref{II}), we have 
 $\mathbbm{1}_{\{|\psi(\sigma(\hat{\mathbf{\eta}}), \mathbf{g})| \geq \alpha \}} = \mathbbm{1}_{\{|\psi(\sigma(\mathbf{\eta}), \mathbf{g})| \geq \alpha \}},$ $\forall \sigma \in \mathcal{S}_n.$

\hspace{-5pt}
\begin{large}
\begin{flushright} 
$\Box$
\end{flushright}
\end{large}


\section{Simulations}\label{sec:sims}

For testing the method, we generated a finite population consisting of 500 clusters with 20 observations each, for a total population size of 10,000 observations.
Each observation $\mathbf{u}_{ij} = (Y_{ij}, A_{ij}, B_{ij}, C_{ij}),$ where $\mathbf{u}_{ij}$ is observation $j$ of cluster $i,$ contains values for 4 random variables.
The continuous random variable $Y$ represents the variable of interest and variables $A$-$C$ the corresponding group labels.

All of the group labels were generated from Bernoulli random variables with equal probability ($.5$) and each have varying amounts of clustering.
Label $A$ was generated from iid Bernoulli random variables with $P(A_{ij} = 1) = 0.5,$ for all $i$ and $j,$ so are independent of cluster label.  
Label $B$ was generated from independent Bernoulli random variables with $P(B_{ij} = 1) = p_i,$ where for each cluster $i,$ $p_i$ was drawn from a U~$(0,1)$ random variable, so each cluster has more or less observations labeled 1 than other clusters.
Labels $C_{ij} = C_i,$ for all $j$ in cluster $i,$ where $C_{i}$ was generated from iid Bernoulli random variables with $P(C_{i} = 1) = 0.5,$ for each $i.$ 
Therefore, every observation has the same label $C$ within a cluster.

\begin{figure}[htb]
\begin{tabular}{c}
\vspace{-.3in}
\includegraphics{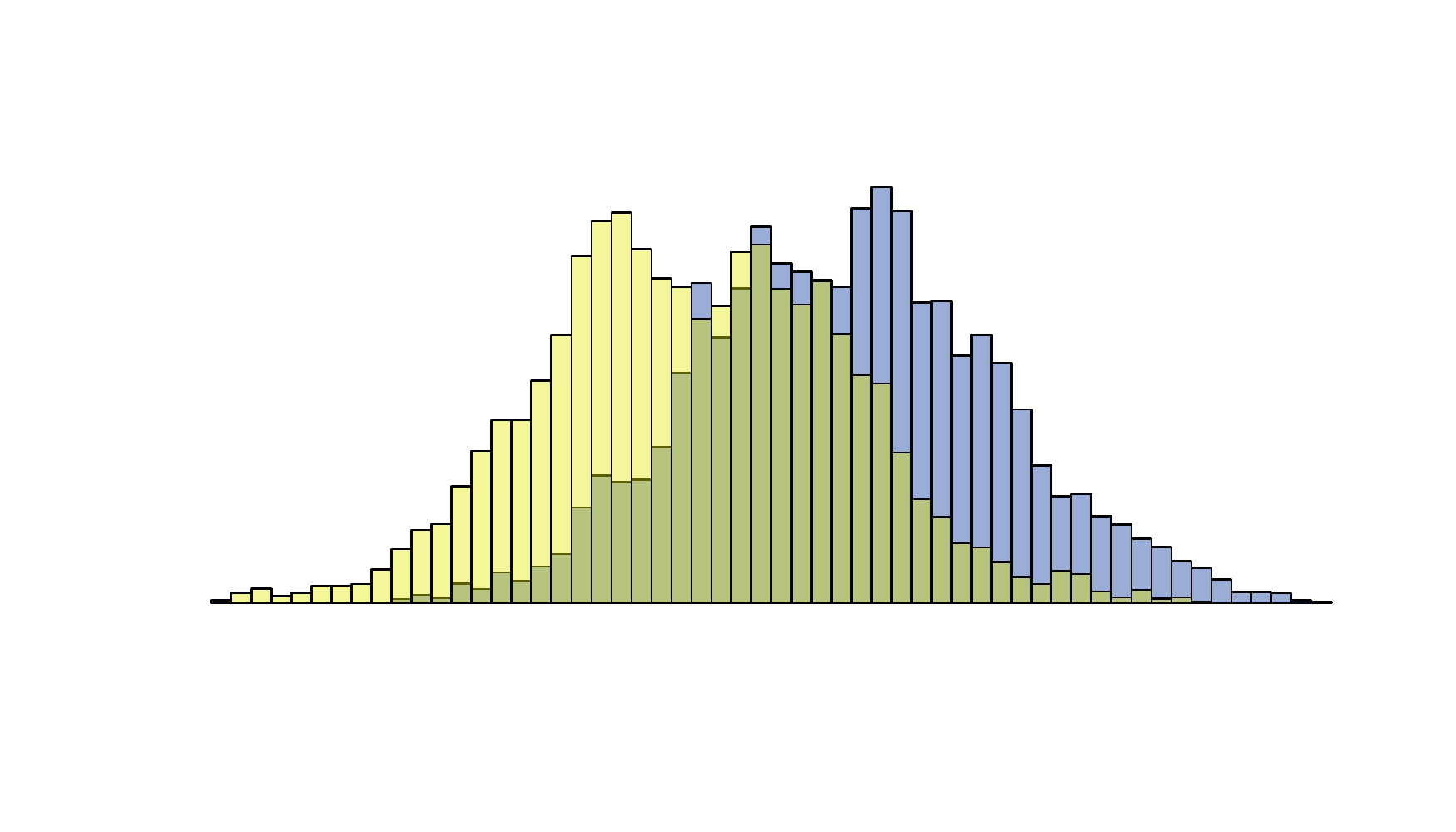}
\end{tabular}
\caption{\footnotesize Distributions of the simulated values for $Y$ in the finite population used for the simulations for group 0 (light color) and group 1 (dark color) when $\delta = \sigma_{\eta}.$ }\label{fig:pop}
\end{figure}

The observations of the variable of interest were generated as iid random variables with distribution given by 
\begin{equation}\label{simy}
Y_{ij} = \mu_{ij} + \nu_i + \epsilon_{ij},
\end{equation} 
where $\nu_i \sim \mathcal{N}(0, \sigma_{\nu}),$ $\epsilon_{ij} \sim \mathcal{N}(0, \sigma_{\epsilon}),$ and $\mu_{ij}$ is a deterministic function of group label $\mu(g_{ij}) =  \delta g_{ij} - \frac{\delta}{2},$ where $\delta$ is constant and $g \in \{A, B, C\}.$
The simulation results presented in this article were obtained using the values $ \sigma_{\epsilon} = 0.5,$ $\sigma_{\nu} = 4,$ and
$\delta = 0$ or $\delta = \sigma_{\eta},$ where $\sigma_\eta$ is the standard deviation of the random variable $\eta,$ defined by equation (\ref{clus_mod}).
Figure \ref{fig:pop} shows the distribution of the simulated values for $Y,$ when  $\delta = \sigma_{\eta}.$


We compared the performance of a hypothesis test based on the proposed pseudo-permutation method to the regular permutation test. 
The test was done over several different sample designs of different sizes.
Taking 2,000 independent samples from the finite population, using a given sample design, and computing the p-value obtained from the proposed test and the regular test for each sample, we obtain a vector of 2,000 estimated p-values for each test and the corresponding value of the test statistic given in equation (\ref{teststat}).

When the null hypothesis is true, $\delta=0,$ the empirical distribution of the set of test statistic values can be used to estimate the true p-value.
This estimate is then compared to the estimated p-values from the permutation tests obtained for each test-static value.
When the null hypothesis is false, $\delta>0,$ the empirical distribution of the set of estimated p-values from a permutation test can be used to estimate the power of the test.

\begin{figure}[htb]
\begin{tabular}{c}
\vspace{-.8 in}
\includegraphics{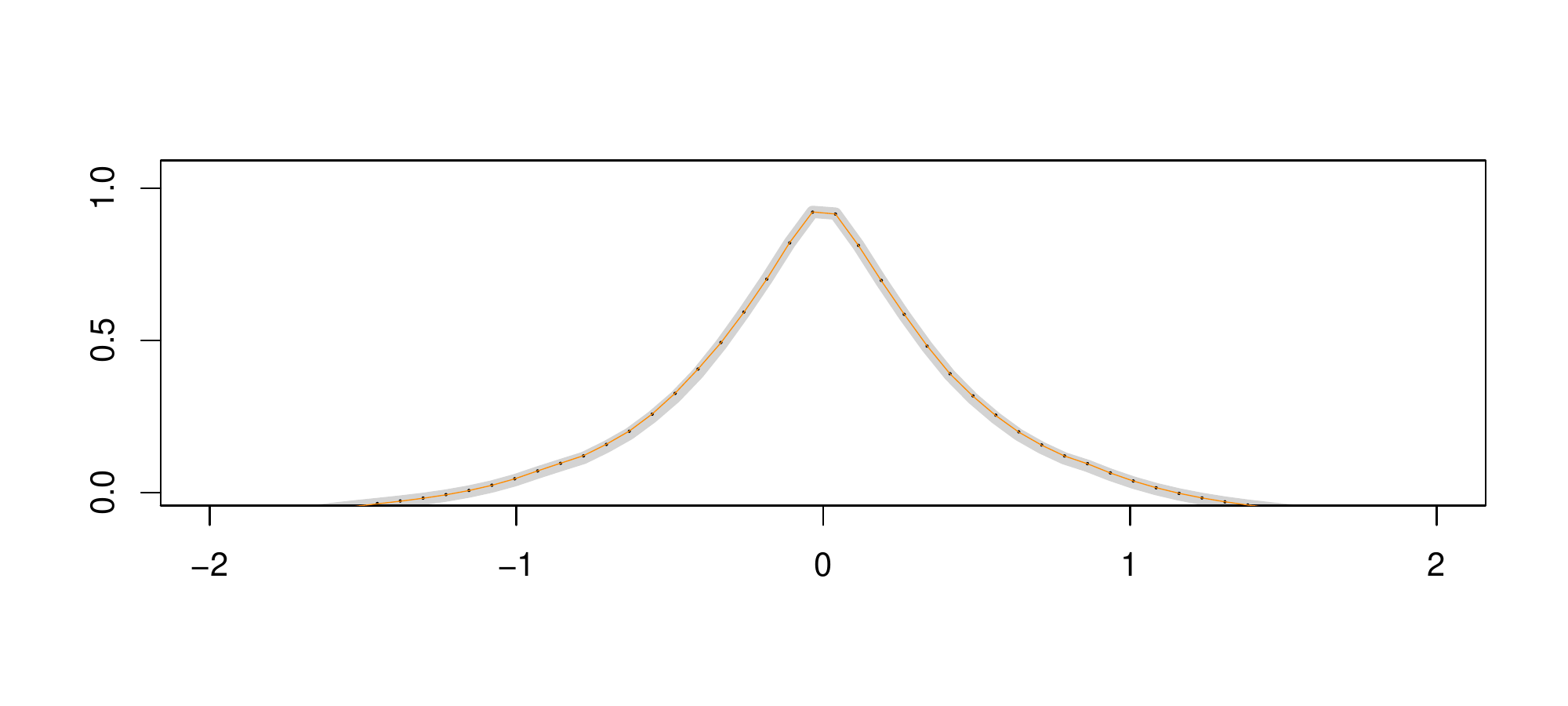}
\\
\includegraphics{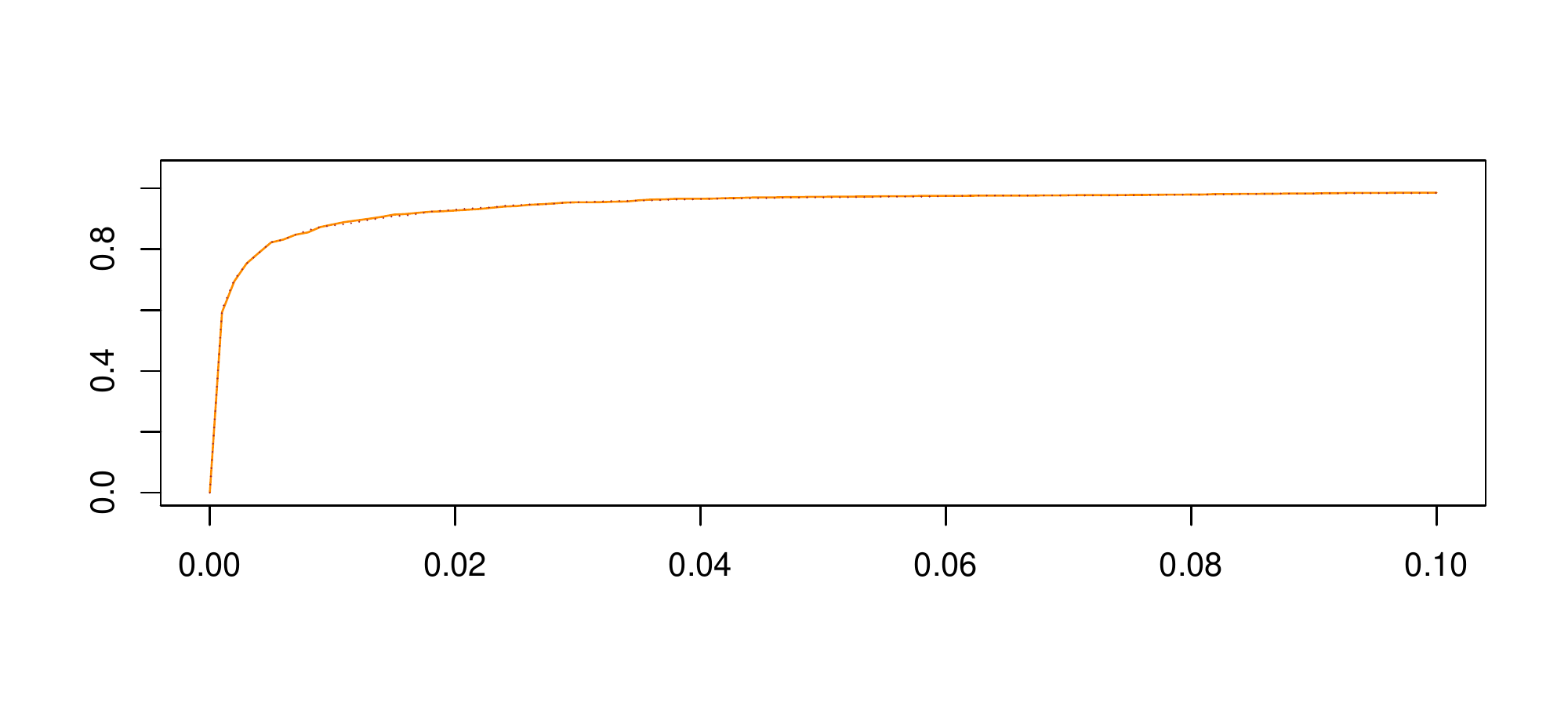}
\end{tabular}
\caption{\footnotesize Top graph displays a comparison of estimated p-values to the true empirical p-values observed over 2,000 simple random samples of size 60 for group labels $B$ when $\delta = 0.$  
Shown are the empirical $p$-values (thick light-grey line), $p$-value calculated from the permutation method ignoring the sample (black dashed-line) and the $p$-value calculated using the pseudo-permutation method accounting for the sample design (orange solid-line) for each value of the test statistic obtained using the sampled data.
The bottom graph displays the power (the proportion of times the test rejected the null) when $\delta = \sigma_{\eta}$ of the regular permutation test (black dotted-line) and the pseudo-permutation test (orange solid-line) for p-values between 0 and 0.1.}\label{fig:srs60}
\end{figure}

For example, consider the test on the label variable $B.$  
The top graph of Figure \ref{fig:srs60} displays (thick light-grey line) the p-values estimated from the empirical distribution of test statistics observed over the 2,000 simple random samples (srs) of size 60 along with the estimated p-values from the pseudo-permutation test (orange solid-line) and the regular permutation test (black dotted-line) over the observed values of the test statistic.
Under the srs design, the regular permutation test gives p-values that match the empirical distribution perfectly; the black dotted-line overlaps the empirical distribution.
The pseudo-permutation test gives higher estimated p-values for lower values of the test statistic than the empirical distribution and regular permutation test, which leads to having less power than the regular test.  

The bottom graph of Figure \ref{fig:srs60} displays the power (the proportion of times the test rejected the null) when $\delta = \sigma_{\eta}$ of the regular permutation test (black dotted-line) and the pseudo-permutation test (orange solid-line) for p-values between 0 and 0.1.
The pseudo-permutation test can be seen to have lower power than the regular test when testing at low (< .02) significance levels under a srs design.

Though the estimated p-values obtained from the pseudo-permutation test a little higher (too high) than the regular test for most values of the test statistic, the regular test provided p-values that are slightly too low. 
Indeed probability of rejection when the null hypothesis is true, is 0.053 for the pseudo-permutation test compared to $0.052$ for the regular test at the .05 level and 0.012 compared to 0.01 at the .01 significance level.
Overall the p-values produced by both tests were acceptable in all of the srs designs (for all variable labels $A$-$C$ and sample sizes $\geq 20$) we tested.  

The performance of both tests improved on data from stratified designs. 
For our tests, we stratified the population based on the quartile values of $\eta,$ group label, or both.
Both tests improved for stratified designs even when the units were sampled with unequal probability of selection, when the probabilities were related to the group label being tested.
When the sample design included unequal probabilities of selection that were related to the values of $Y,$ only the pseudo-permutation test (adjusted for the sample design) performed reasonably.

\begin{figure}[htb]
\begin{tabular}{c}
\vspace{-.8in}
\includegraphics{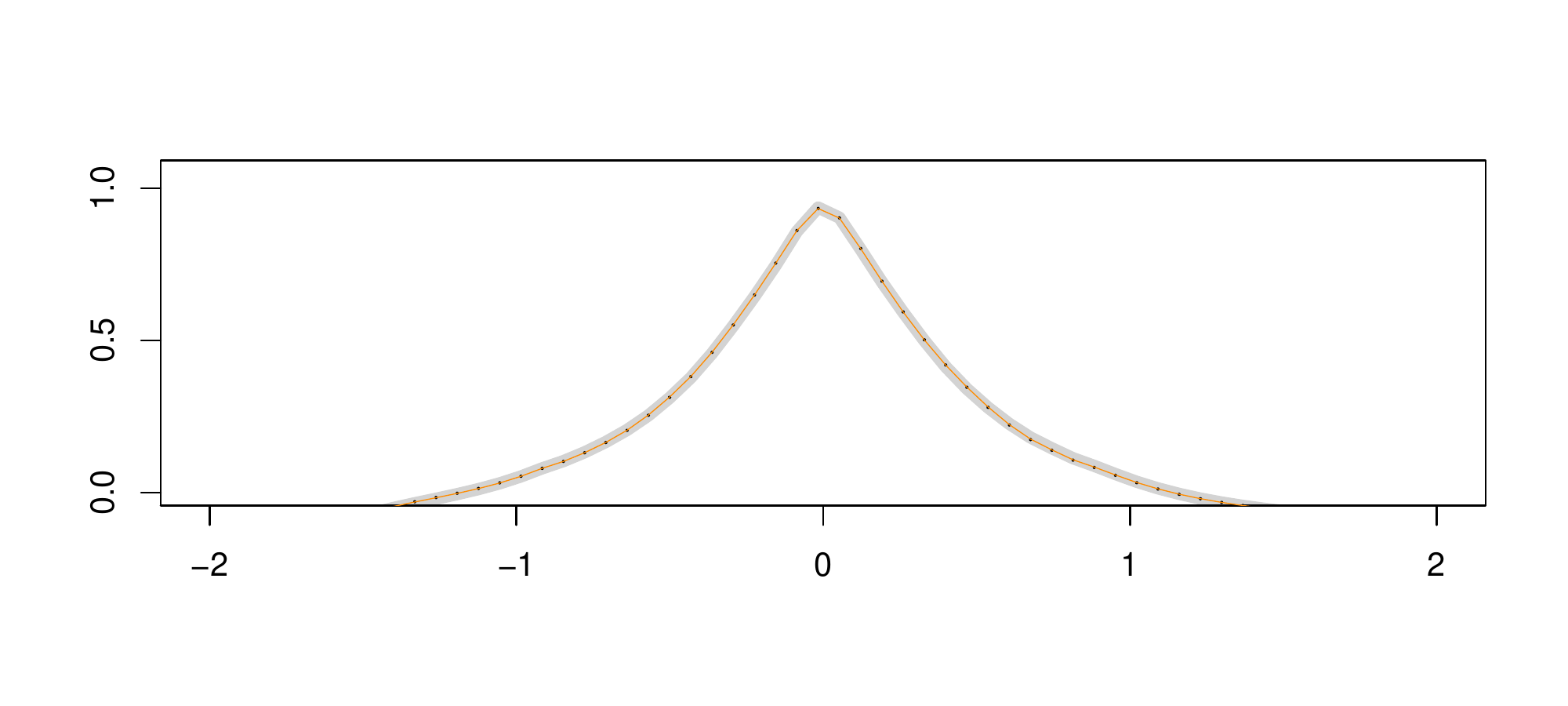}
\\
\vspace{-.8in}
\includegraphics{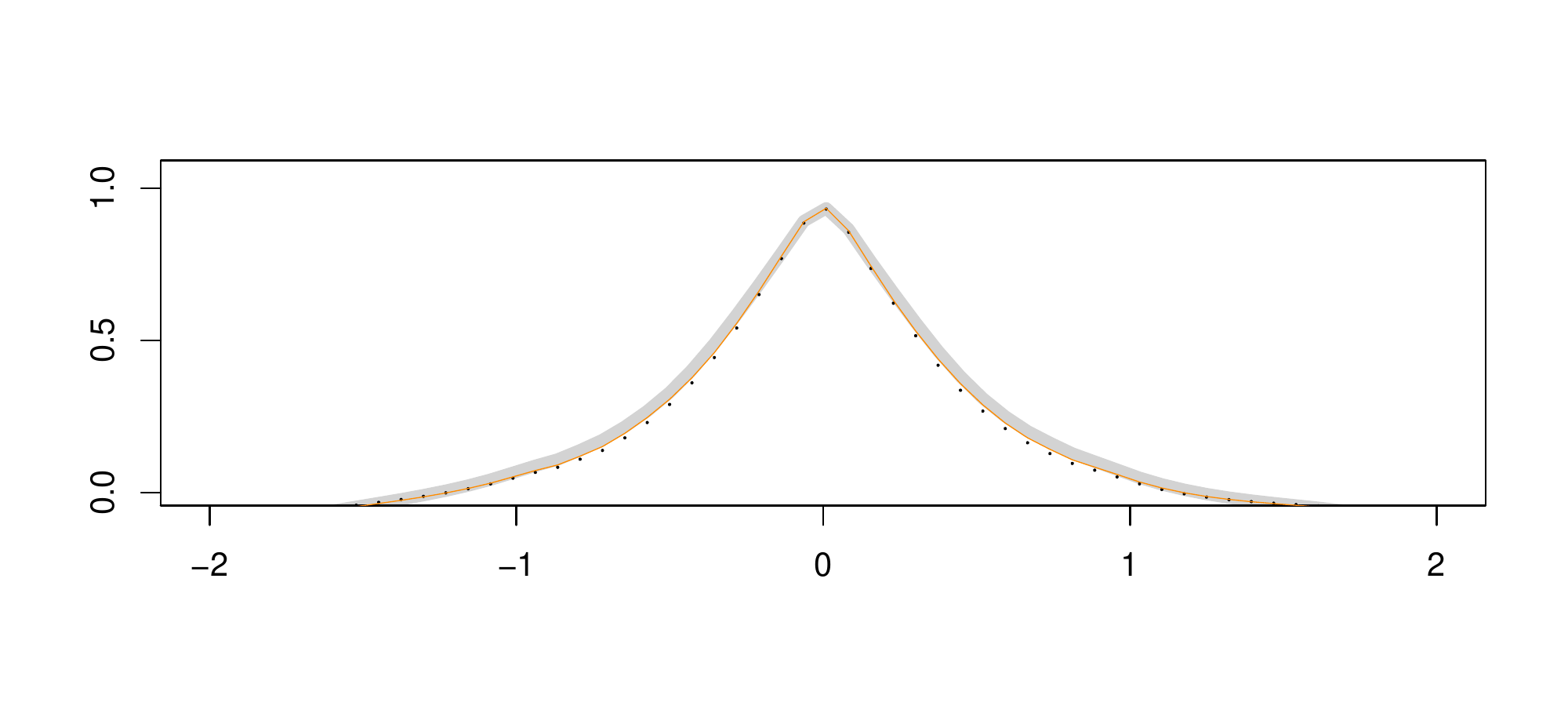}
\\
\vspace{-.8in}
\includegraphics{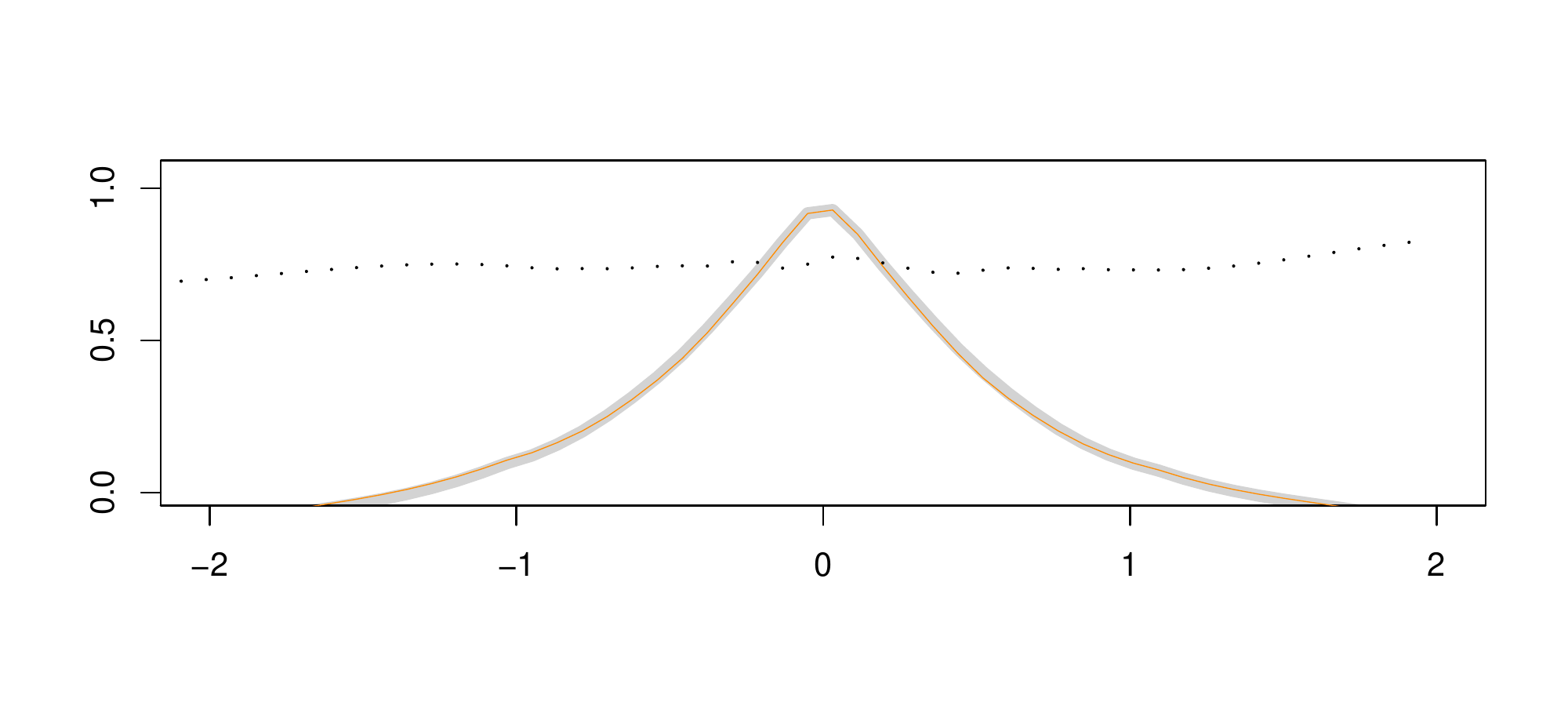}
\\
\vspace{-.3in}
\includegraphics{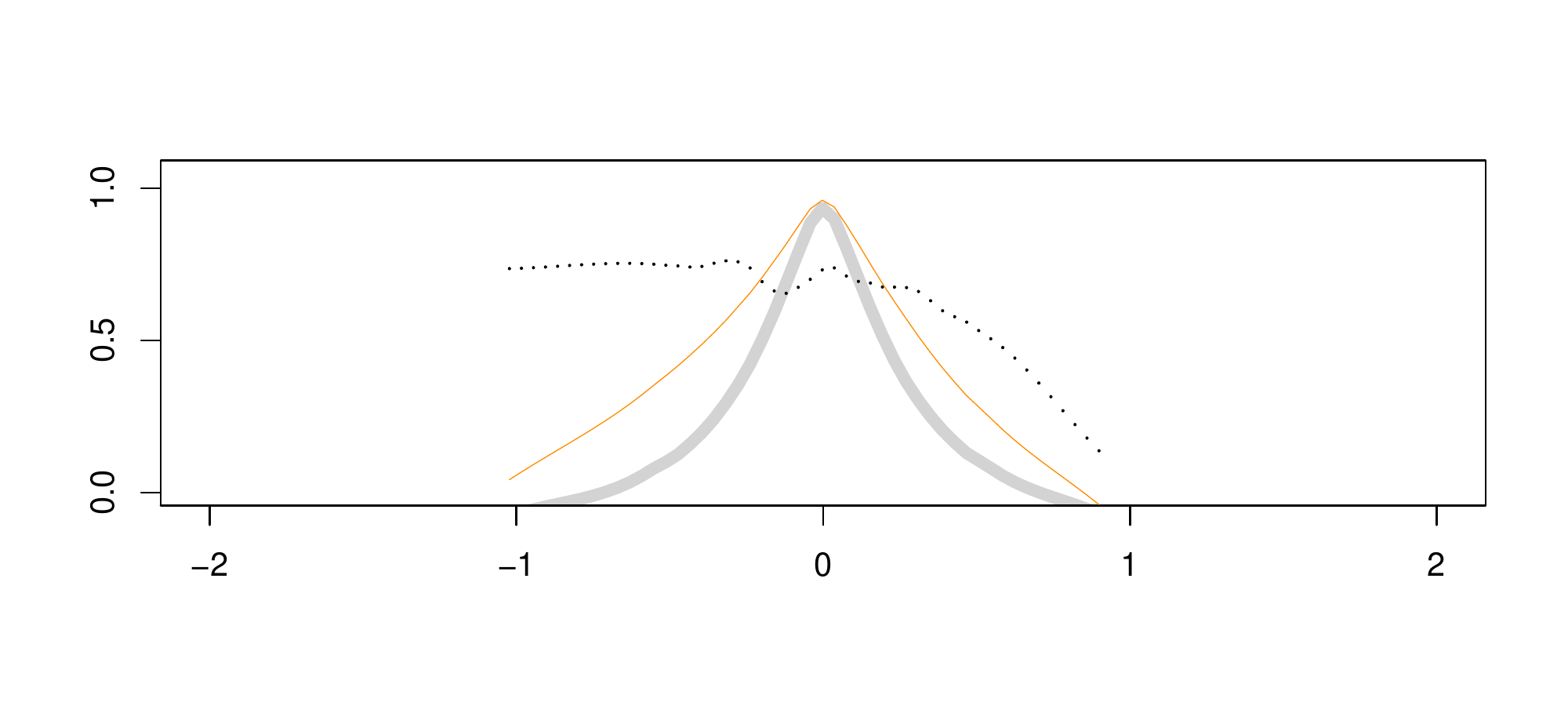}
\end{tabular}
\caption{\footnotesize Test on data with $\delta = 0$ over 2000 repeated samples of size 60 for various unequal weighted samples.
Shown are the empirical $p$-values (solid light-grey line), $p$-value calculated from the permutation method ignoring the sample (black dashed-line) and the $p$-value calculated using the method accounting for the sample design (orange solid-line) for each value of the test statistic obtained using the sampled data.}\label{fig:nullother}
\end{figure}

Figure \ref{fig:nullother} shows the results of the tests on label $B$ when $\delta=0$ for stratified sample designs.
The results in the top two graphs respectively are for a stratified equal probability of selection design and a stratified design, where units with group label $B = 1$ were sampled at twice the rate as units with group label $B = 0.$
The estimated p-values of both tests follow the empirical p-values under both designs. 

The third graph in Figure \ref{fig:nullother} contains results of the test for a stratified design where the probability of selection was higher for larger quartiles.
In this case, the test using the weight adjusted estimator produces p-values that closely follow the empirical p-values, while the unadjusted test failed to produce reasonable p-values.
For example, the probability of mistakenly rejecting the null-hypothesis using a 5\% confidence level was only 0.004 for the unadjusted test compared to 0.05 for the pseudo-permutation test adjusted using the sample weights.

The final graph in Figure \ref{fig:nullother} shows the results for a design with strata based on quartiles of $\eta$ and group label $B.$
Units were selected so that units in the larger quartiles of $\eta$ and with label $B=1$ were selected with higher probability than units in lower quartiles of $\eta$ or with label $B=0.$
In this case, the varying weights made the tests less efficient, but again only the pseudo-permutation test adjusted using the sample weights produced reasonable p-values estimates.

\begin{figure}[htb]
\begin{tabular}{c}
\vspace{-.8 in}
\includegraphics{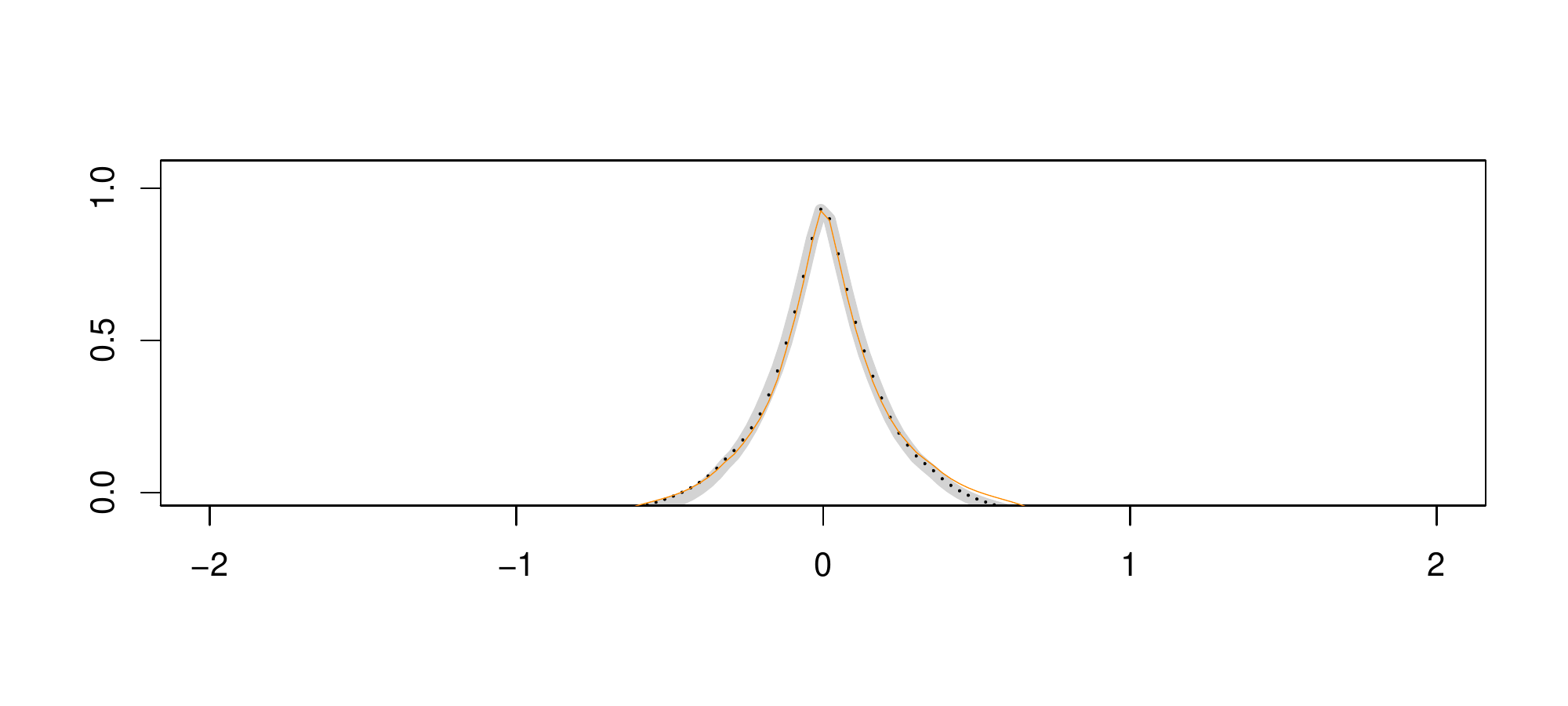}
\\
\vspace{-.8 in}
\includegraphics{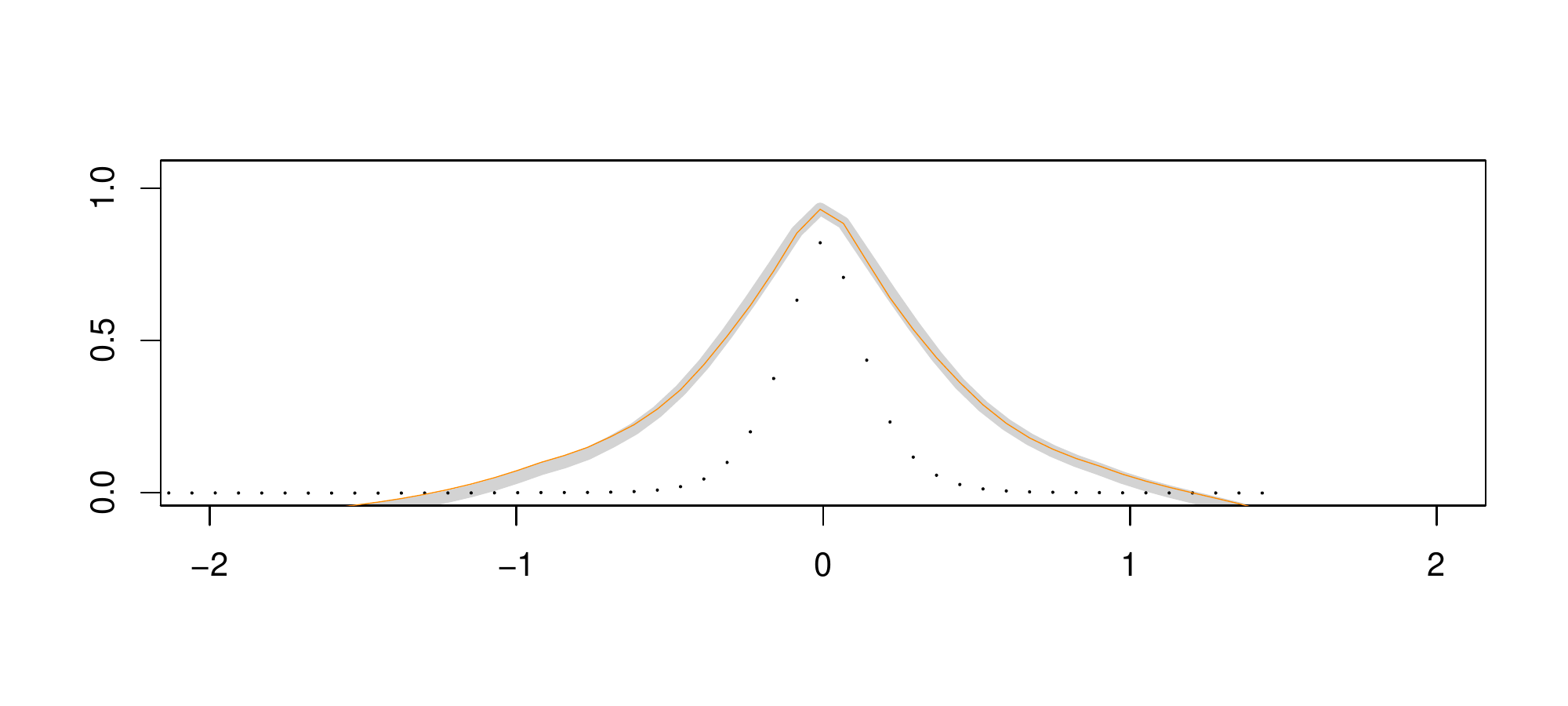}
\\
\vspace{-.3 in}
\includegraphics{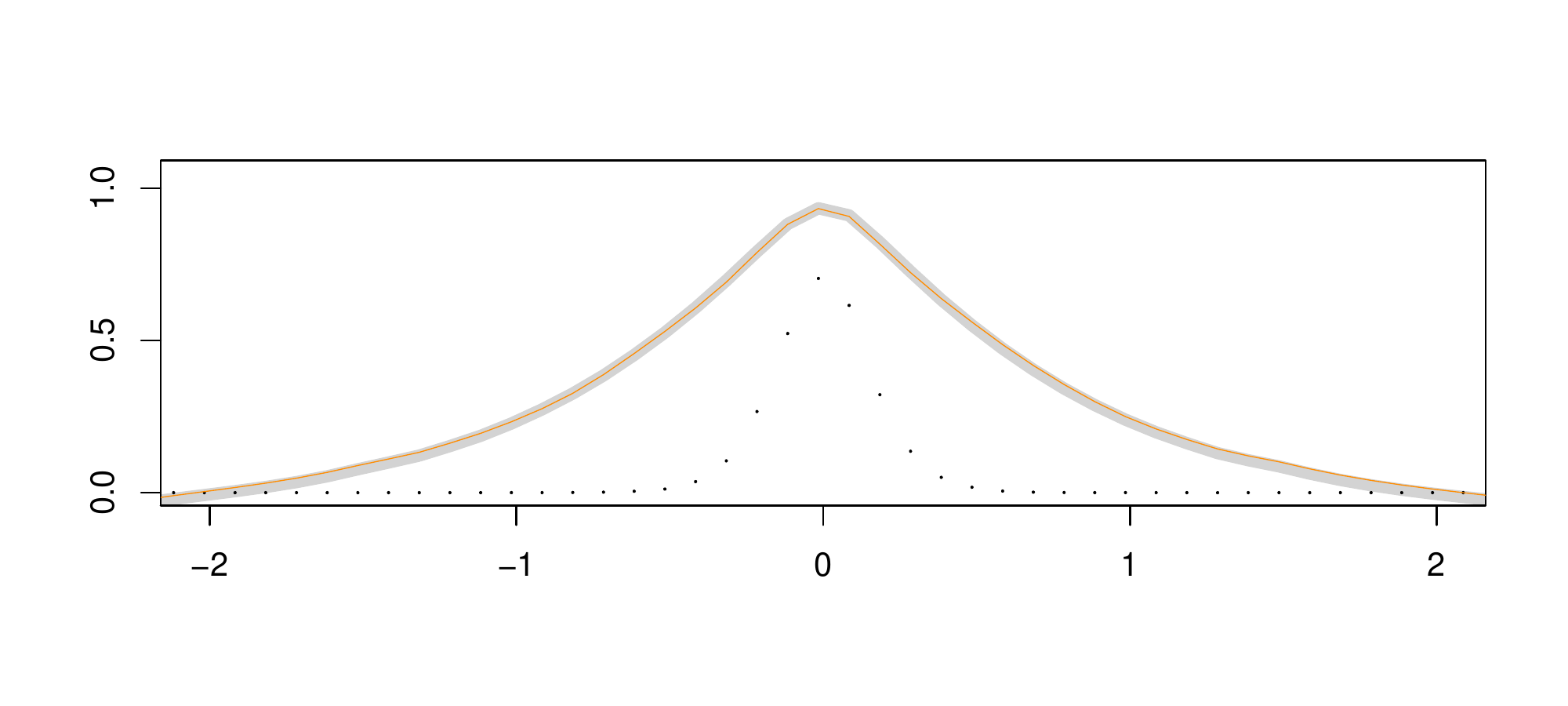}
\end{tabular}
\caption{\footnotesize Test on data with $\delta = 0$ over 2,000 repeated samples of 20 randomly selected clusters.
Shown are the estimated $p$-values from the empirical distribution (thick light-grey line), from the permutation method ignoring the sample (black dotted-line) and the method accounting for the sample design (orange solid-line) for each value of the test statistic obtained using the sampled data.
The three graphs (from top to bottom) display the test results for group labels $A,$ $B,$ and labels $C.$}\label{fig:nullcs20}
\end{figure}

Figure \ref{fig:nullcs20} displays the results of the tests on 2,000 repeated samples of 20 randomly selected clusters,
when $\delta = 0$ over group labels $A,$ $B,$ and $C.$
The top graph displays the results for the test of group label $A$ under the null-hypothesis.
In this case, the estimated p-values from the two-tests and the empirical distribution are indistinguishable because the cluster ids and the labels are independent; thus both tests in this case do an excellent job providing approximations to the true p-value of the test statistic.
The results displayed in the bottom two graphs of Figure \ref{fig:nullcs20}, testing group labels $B$ and $C$ respectively, demonstrate that ignoring the cluster design when the group labels are more homogeneous within cluster, leads to misleading low p-value estimates using the regular permutation test.
Meanwhile, the proposed pseudo-permutation test produces p-values that closely match the empirical distribution under all three labels.

\begin{figure}[htb]
\begin{tabular}{c}
\vspace{-.8 in}
\includegraphics{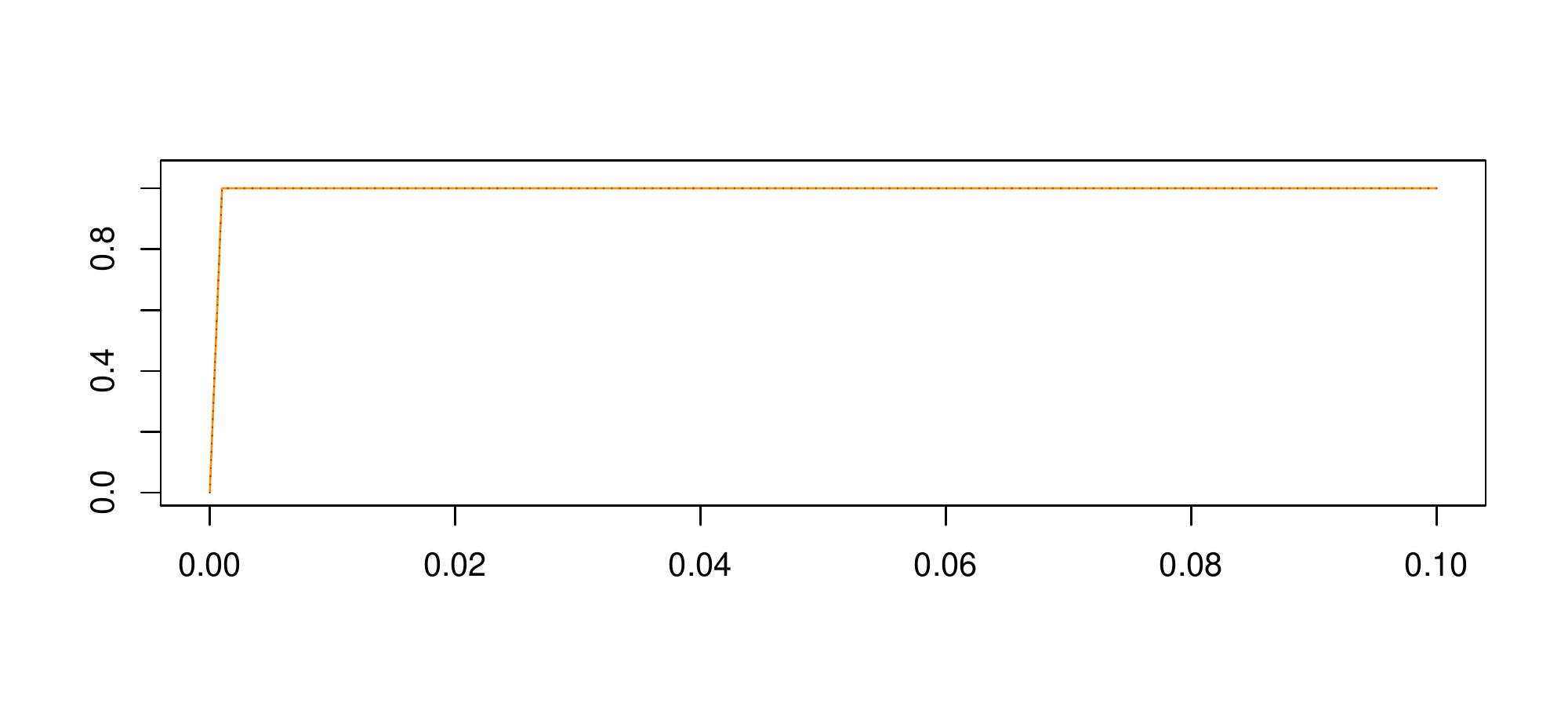}
\\
\vspace{-.8 in}
\includegraphics{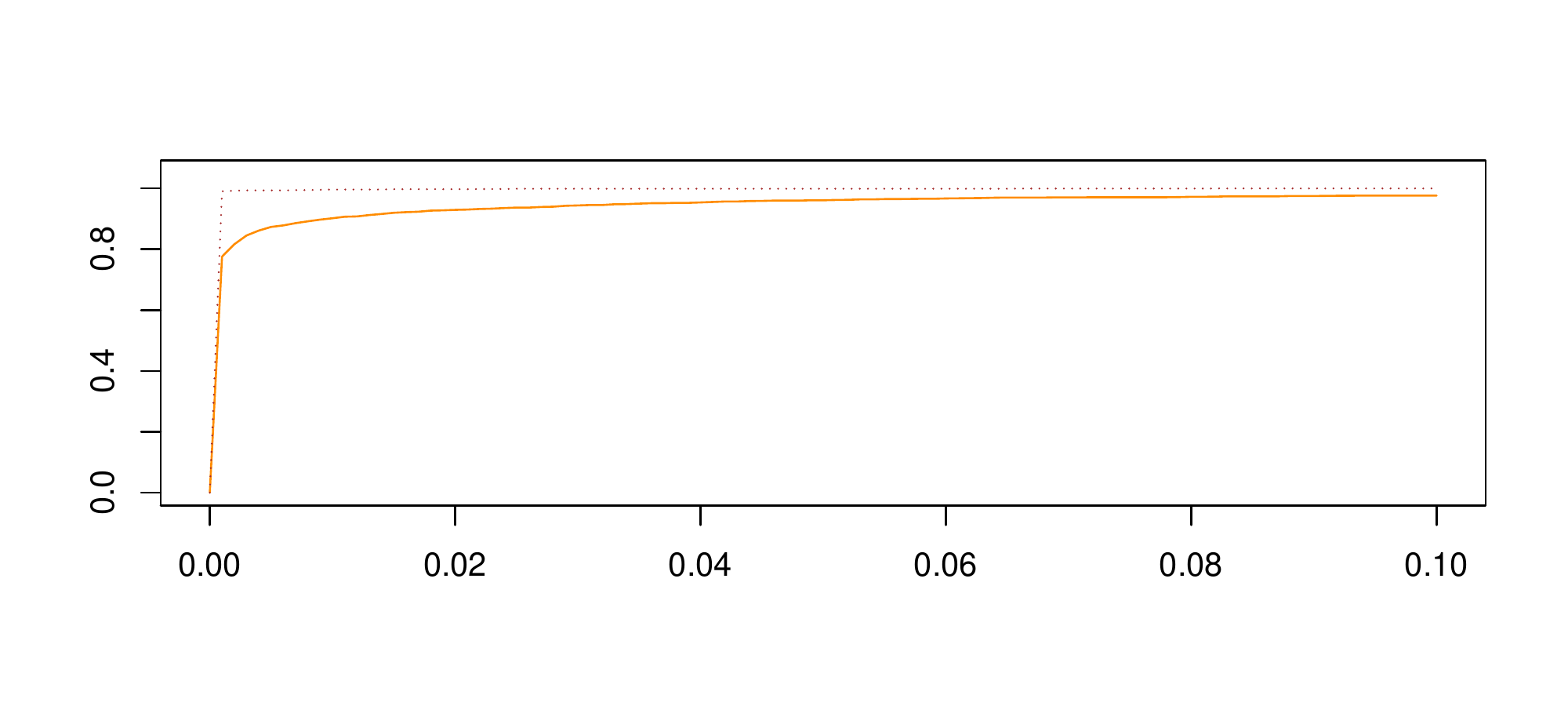}
\\
\vspace{-.3 in}
\includegraphics{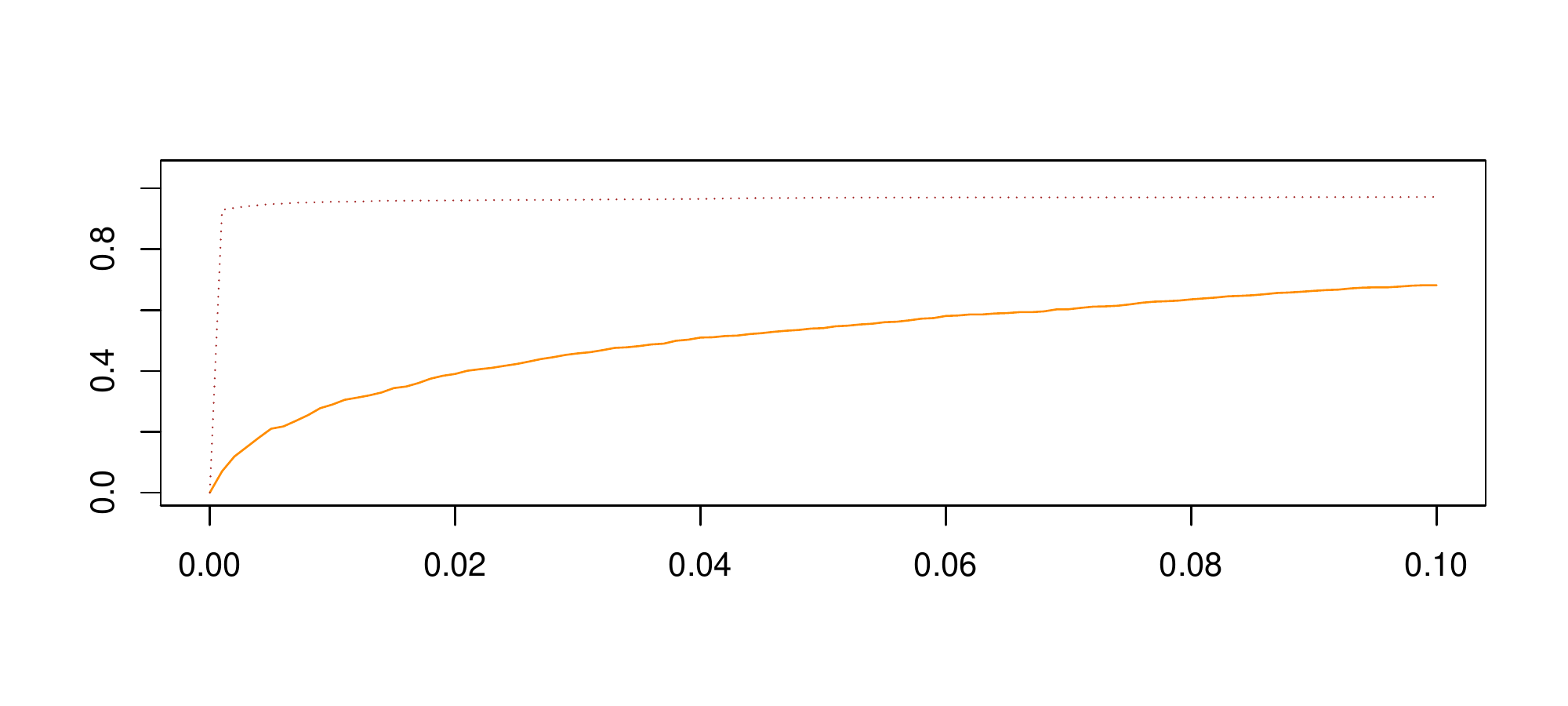}
\\
\end{tabular}
\caption{\footnotesize ROC curve with $P\big(\mbox{reject null})$ on the $y$-axis and $p$-value on the $x$-axis.  
        This is for the data from the test when the $\delta = \sigma_{\eta}$ over 2,000 repeated samples of 20 randomly selected clusters }\label{fig:20cs1a}
\end{figure}

This robustness of the pseudo-permutation test under cluster sampling comes at a cost in power over the regular permutation test when there is no association between the group labels and cluster id.
The three graphs of Figure \ref{fig:20cs1a} show the power of the pseudo-permutation test (orange solid-line) and the regular permutation test (red dotted-line) over different levels of significance when $\delta = \sigma_{\eta},$ for the three group labels: $A,$ $B,$ and $C$ respectively, when the data comes from a simple random sample of 20 clusters.
We can see that two tests have the same power testing group $A,$ but the pseudo-permutation test has considerably less power than the regular test when testing groups $B$ and $C.$
Since the regular permutation test produces p-values that are much too small under the null hypothesis for tests on labels $B$ and $C,$ the power is meaningless for this test and only provided for reference to compare against the power of the pseudo-test.

Looking at the last graph of Figure \ref{fig:20cs1a}, the pseudo-permutation test for label $C$ in particular appears to has considerably reduced power.
However, because all units in each cluster have the same label value, the pseudo-permutation test reduces to a test of 20 observations.
In fact, the two-sided t-test on 20 observations at 5\% significance level has power 0.562 compared to 0.54, the power of the pseudo-permutation test.

\section{Consumer Expenditure Survey Data}\label{sec:app}




For the illustration of this method we use data from the U.S. Bureau of Labor Statistics Consumer Expenditure (CE) survey to test differences in earnings and spending between families with a primary earner that has at least a bachelor's degree against families with a primary earner that does not.
We will refer to the groups as the "college educated" group ($g=1$) and the "not college educated" group ($g=0$) respectively.
A subset of variables contained in the CE 2015 interview dataset are provided in the \texttt{rpms} package \citep{rpms}.
This dataset includes variables on the sample design, household, and person listed as the household's primary earner.

We test for differences between these groups on two quantitative variables (household income and family size) and two proportions (proportion of families that have expenditures on tobacco and proportion of families that have a vehicle).
For our analysis we will consider households with primary earners between the ages of 22 and 64 and where the education level of the primary earner, household income and expenditure information is provided.
This gives us a sample size of 50,762 from 115 sampled clusters.
Table \ref{tab:ce} shows the comparisons between the college educated group and non college educated group as well as the estimated $p$-value from the permutation test ignoring clusters and the permutation test adjusted for the cluster membership.
In both tests, we first subtracted the estimated unconditional mean of each random variable using the sample weights and performed the test on the residuals.



\begin{table}[ht]
\centering
\begin{tabular}{|p{130pt}|c|c|c|c|}
  \hline
  \multicolumn{1}{|c}{\multirow{2}{110pt}{\textbf{Description}} } & \multicolumn{2}{|c|}{\textbf{College Educated}} & \multicolumn{2}{c|}{\textbf{Estimated p-value}} \\ 
  \cline{2-5}
   & \textbf{Yes}  & \textbf{No} & \textbf{iid} & \textbf{cluster}\\ 
   \hline
   \hline

Sample Size &  18,175 & 32,587 & - & - \\ \hline

Mean Household Income &  94,584  & 46,487 & 0 & 0\\ \hline

Mean Family Size &  2.5795  & 2.7888 & 0 & 0.313\\ \hline

Proportion with a Vehicle & 0.9329 &  0.8705 & 0 & 0.3905 \\
\hline

Proportion of Using \newline Tobacco & 0.0609 &  0.1868 & 0 & 0 \\ 
\hline

%

\end{tabular}\caption{CE 2015 Comparison of families that have a primary earner with at least a 4-year college degree to families that have a primary earner that does not have a 4-year degree.}\label{tab:ce}

\end{table}

The two continuous random variables, family size and household income before tax, are both available on the CE dataset.
In order to estimate the proportion of families with vehicles, we made an indicator random variable equal to 1 if the sum of the reported number of vehicles owned and the vehicles leased was greater than 0.
Similary, for estimating the proportion of families using tobacco, we made an indicator random variable that is equal to 1 if the reported expenditure on tobacco was greater than 0.

It is interesting to note that when we treat all the observations as independent (ignoring clustering), the permutation test finds the difference between every variable considered significant.  
However, when we accounted for the clustering, the difference in family size and the proportion of families with a vehicle was not found to be significant.
These results seem to be reasonable as we would expect household income for families where the primary earner has a college degree to be larger than those with a primary earner without a degree.
Likewise, it is probable that the more educated families would be less likely to use tobacco due to the many scientific reports linking tobacco use to a variety of health issues, but it is not clear that they would be more or less likely to have a car or have a larger or smaller family size.

These comparisons are intended to illustrate the method, and the results are encouraging in that they seem to highlight the importance of adjusting for the sample design when using sample data.  
Because the simulation results show that this method leads to a loss of power compared to the iid test, we cannot be sure that there is not a difference between family size and proportion of families with a car between these two groups, but they also show that ignoring the sample design is likely to lead to completely unreliable estimated $p$-values.
Since the variables tested are likely to be correlated within clusters, we would not trust any results ignoring the sample design.

\section{Discussion}\label{sec:disc}

We have proposed a general method for performing a pseudo-permutation test that accounts for the complex sample design and have shown that the test will give design consistent results under a set of conditions on the sample design and population structure.
Tests using a simulated population comparing the performance of the proposed method to permutation tests that ignore the sample design demonstrate that it is important to account for the sample design in order to obtain reasonable $p$-value estimates.
The results of these simulations and an application using publicly available consumer expenditure data especially highlight the importance of accounting for clustering in the sample.

Though accounting for the sample design protects against performing an invalid test when the design is informative, the presented permutation method also leads to a loss of power.   
This loss of power occurs whether the sample design is informative with respect to the variable of interest or not. 
Perhaps, this could be mitigated by adjusting the proposed method using an estimate of the design-effect in some way, which could be the subject of further research.
However, the presented method represents a general method for performing a permutation test on data obtained through a complex sample that will provide valid inference at the cost of some power.   

\paragraph{Acknowledgments\\}

The authors would like to thank people (specifics to be added later).

\bibliographystyle{plainnat}

\bibliography{perm_refs}{}

\clearpage\pagebreak\newpage \thispagestyle{empty}

\end{document}